\documentclass[runningheads]{llncs}

\usepackage[T1]{fontenc}
\usepackage{graphicx}
\usepackage{longtable}   
\usepackage{booktabs}    
\usepackage{caption}     
\usepackage{array}       
\usepackage{tcolorbox}
\usepackage{subcaption} 
\usepackage{url}

\begin{document}
\title{Organizational Cohesion in Microservice Architectures: A Multi-Project Empirical Study}
%
\titlerunning{Organizational Cohesion in Microservices}
%

\author{
Xiaozhou Li\orcidID{0000-0002-3767-2527}
\and
Andrea Janes\orcidID{0000-0002-1423-6773}
}
\authorrunning{X. Li and A. Janes}
\institute{
Free University of Bozen-Bolzano,\\ via Bruno Buozzi, 1, 39100,\\Bolzano, Italy\\
\email{\{xiaozhou.li, andrea.janes\}@unibz.it}\\
}

\maketitle 
\begin{abstract}

The widespread adoption of microservice architectures has introduced new challenges in aligning software modularity with the structure of development organizations. Although prior research has extensively examined technical properties such as service coupling and dependency structures, comparatively little attention has been paid to how contributor activity reflects or diverges from service boundaries. In this paper, we introduce the notion of organizational cohesion in microservice ecosystems and propose a quantitative approach to measure it. Building on the Sensitive Class Cohesion Metric (SCOM), we define Pairwise Team Cohesion (PTC), a metric that captures the balance and focus of developer contributions within individual microservices. We analyze the evolution of organizational cohesion using a longitudinal case study of the Spinnaker microservice platform and replicate the analysis across six additional open-source microservice systems. Our results reveal systematic differences between core and peripheral services and show that PTC and Average Organizational Coupling (AOC) exhibit only a weak correlation across projects. This finding shows that team cohesion and cross-service developer activity suggest distinct and weakly associated organizational dynamics. By extending the "high cohesion, low coupling" principle to the organizational level, our study provides a quantitative perspective for assessing the socio-technical structure of microservice development.

\end{abstract}

\section{Introduction}
\label{sec:introduction}

Microservice architecture has become a dominant paradigm for building large-scale cloud-native software systems \cite{dragoni2017microservices}. It decomposes the application into a set of independently deployable services that encapsulate specific business capabilities and communicate through lightweight APIs \cite{newman2021building}. 
However, this architectural style also introduces substantial organizational complexity when development is distributed across multiple teams whose coordination must align with service boundaries \cite{lewis2014microservices}. Hence, the effectiveness of a microservice system depends not only on its technical modularity, but also on how well the development organization reflects and sustains that modular structure.


The relationship between software architecture and organizational structure has long been recognized through Conway’s Law, which posits that system design mirrors communication structures \cite{conway1968committees}. Especially for microservices, this implies that each service ideally corresponds to a cohesive team with clear ownership; however, such alignment is often imperfect \cite{amoroso2023one}. Therein, developers contribute across multiple services, responsibilities overlap, and coordination patterns evolve over time. Understanding how contributor activity aligns with service boundaries is therefore essential for assessing the socio-technical health of microservice ecosystems.


Cohesion and coupling are long-established principles of software design. Cohesion refers to the degree to which elements within a module belong together, while coupling measures the degree of interdependence between modules \cite{yourdon1979structured}. High cohesion and low coupling are widely associated with maintainability and architectural quality \cite{parnas1972criteria}. Although extensive research has operationalized these concepts for source code, comparatively little work has quantified them at the organizational level. In particular, while organizational coupling has been studied through cross-service contribution patterns \cite{li2023evaluating}, there is a lack of systematic approaches to measure the cohesion of microservice teams.

To address this gap, we introduce Pairwise Team Cohesion (PTC), a quantitative metric that captures the extent to which contributors to a microservice exhibit focused and balanced participation. PTC is derived from version-control data and adapts principles from established cohesion metrics to the organizational context by modeling the overlap and distribution of developer activity. Complementing this, we use Average Organizational Coupling (AOC) to quantify cross-service developer overlap, providing a measure of inter-team interdependence. Importantly, we treat commit activity as a scalable behavioral proxy for contribution patterns, while acknowledging that it does not fully capture communication or coordination processes. This study addresses the following research questions: 

\begin{description}
\item[RQ1.] \textit{How can the cohesion of microservice teams be quantitatively evaluated using development activity data?}
\item[RQ2.] \textit{What is the relationship between organizational cohesion and organizational coupling in microservice systems?}
\end{description}



    
    
    

To answer these questions, we conduct a longitudinal case study of the \textit{Spinnaker}\footnote{https://spinnaker.io/} microservice platform and complement it with a cross-system replication across multiple open-source microservice projects. We also performed a robustness analysis to assess the sensitivity of the results to data perturbations, contributor effects, metric formulations, and temporal aggregation strategies.

The results show that PTC captures meaningful patterns of contributor focus and responds systematically to organizational and architectural changes. More importantly, across systems and experimental conditions, the correlation between cohesion and coupling remains consistently weak. This indicates that cohesion and coupling provide distinct, weakly associated, and non-redundant views
of organizational structure, rather than opposing ends of a single spectrum. This finding reframes the classical "high cohesion, low coupling" principle at the organizational level as a two-dimensional design space.

This study makes three main contributions: 1) PTC as a reproducible metric for contribution-based organizational cohesion in microservice systems, 2) empirical evidence that PTC and service-level organizational coupling capture distinct and weakly associated dimensions of developer behavior, and 3) a practical framework to analyze team–architecture alignment using version-control data.

The remainder of this article is organized as follows. Section~\ref{sec:relatedwork} reviews related work. Section~\ref{sec:methods} presents the proposed metrics for organizational cohesion and coupling. Section~\ref{sec:studydesign} describes the study design. Section~\ref{sec:results} reports the empirical results, first summarizing the longitudinal case study of Spinnaker and then presenting the cross-system replication as the primary evidence. Section~\ref{sec:robustness} evaluates the robustness of the findings. Section~\ref{sec:discussion} discusses the implications and limitations of the study. Section~\ref{sec:threat} outlines threats to validity, and Section~\ref{sec:conclusion} concludes the article.
\section{Related Work}
\label{sec:relatedwork}


Cohesion and coupling are foundational concepts in software modularity. Cohesion refers to the degree to which elements within a module belong together, while coupling captures the level of interdependence between modules \cite{stevens1974structured}. Classical work in software design emphasizes that high cohesion and low coupling lead to improved maintainability, understandability, and evolvability. Numerous metrics have been proposed to quantify cohesion at the code level, including variants of LCOM \cite{chidamber1991towards,chidamber1994metrics,li1993maintenance,hitz1995measuring,henderson1995object} and similarity-based measures that assess relationships between methods and attributes \cite{counsell2006interpretation,al2012precise}. Among these, the Sensitive Class Cohesion Metric (SCOM) introduces a weighted formulation that accounts for both the intensity and the distribution of internal relationships, providing a stable and discriminative measure of cohesion \cite{fernandez2006sensitive}.

Although these metrics are well-established for code structure, their application has largely remained within the technical domain. In microservice architectures, cohesion and coupling have also been studied at the service level, focusing primarily on structural dependencies, API interactions, and service evolution \cite{shim2008design,moreira2022analysis}. However, these approaches do not directly capture how developer activity aligns with service boundaries, leaving the organizational dimension of cohesion underexplored.

On the other hand, regarding the organizational perspective of microservice architecture, it is commonly assumed that the alignment between team and service reflects Conway's Law \cite{conway1968committees}. Therefore, ideally, each microservice is typically owned by a semi-autonomous team responsible for its lifecycle, from development and deployment to monitoring and maintenance \cite{newman2021building}. In practice, however, empirical studies show that such alignment is often partial: developers frequently contribute across multiple services, and ownership structures evolve over time \cite{amoroso2023one}. Previous work in microservice organizations has focused primarily on organizational coupling, measured through cross-service contribution patterns \cite{li2023evaluating,li2025exploring}. Recent role-centered evidence further shows that organizational coupling can be shaped by specific developer roles rather than by architectural structure alone \cite{li2026keydeveloperroles}. However, there is limited work on quantifying organizational cohesion, that is, the extent to which contributors focus their activity within a service.

Meanwhile, the relationship between software architecture and team organization has been extensively studied in socio-technical research. Socio-technical congruence (STC) theory argues that effective software development occurs when coordination among developers matches the technical dependencies of the system \cite{cataldo2008socio}. The misalignment between social and technical structures can lead to increased coordination costs, reduced productivity, and quality issues \cite{herbsleb1999splitting}. In this context, cohesion and coupling can be interpreted as complementary organizational dimensions. Cohesion reflects the internal alignment of contributors within a service, while coupling captures cross-service coordination requirements. From a coordination perspective, high cohesion may indicate concentrated ownership and reduced internal coordination overhead, whereas high coupling reflects increased coordination across teams. However, existing STC-based studies primarily focus on communication networks and dependency structures, rather than providing direct quantitative measures of team-level cohesion.



To summarize, previous work provides strong foundations for understanding modularity and socio-technical alignment, but leaves a gap in measuring organizational cohesion in microservice systems. Existing metrics focus either on code structure or cross-service coupling, without capturing the distribution and balance of contributor activity within services. Moreover, the relationship between cohesion and coupling at the organizational level remains largely unexplored. 

\section{Method}
\label{sec:methods}

\subsection{Measuring Organizational Cohesion}
\label{subsec:ptc}

To quantify organizational cohesion in microservice teams, we adapt the Sensitive Class Cohesion Metric (SCOM) \cite{fernandez2006sensitive} to the context of developer activity. The SCOM metric yields a discriminative measure of cohesion, which is robust to minor structure variations and sensitive to balanced participation among elements. While it evaluates relationships among methods based on shared attributes, we reinterpret its core idea at the organizational level: \textit{cohesion arises when contributors focus their efforts consistently within a service and exhibit balanced participation}.

We begin by defining two fundamental components for each developer $d$ and service $s$:

\begin{itemize}
    \item \textbf{Contribution focus}:
    \begin{equation}
    f_{d,s} = \frac{C_{d,s}}{C_{d,*}}
    \end{equation}
    \item \textbf{Contribution weight}:
    \begin{equation}
    w_{d,s} = \frac{C_{d,s}}{C_{*,s}}
    \end{equation}
\end{itemize}

where $C_{d,s}$ is the number of commits made by the developer $d$ to service $s$, $C_{d,*}$ is the total commits by $d$, and $C_{*,s}$ is the total commits to service $s$.



To capture the contributors' joint effort concentration, we introduce Pairwise Team Cohesion (PTC). PTC extends SCOM's pairwise formulation to model cohesion as a shared focus among contributors. For each pair of contributors $(i, j)$ to the service $s$, we define:


\begin{itemize}
    
    \item \textbf{Connection intensity}: 
    \begin{equation}
    C^{(s)}_{ij} = \sqrt{f_{i,s} \cdot f_{j,s}}
    \end{equation}
    
    \item \textbf{Pairwise prominence weight}: 
    \begin{equation}
    \alpha^{(s)}_{ij} = \frac{w_{i,s} + w_{j,s}}{2}
    \end{equation}
    
\end{itemize}

Let $m$ be the number of contributors to service $s$. The Pairwise Team Cohesion is then defined as:

\begin{equation}
PTC_s = \frac{2}{m(m-1)} \sum_{i<j} C^{(s)}_{ij} \cdot \alpha^{(s)}_{ij}
\end{equation}

Regarding the \textbf{Design Rationale}, the formulation of PTC follows two key principles. First, the geometric mean in $C^{(s)}_{ij}$ emphasizes a \textit{balanced focus} between contributors. If one developer is highly focused while another is not, the resulting value decreases, penalizing such an imbalance. Thus, high cohesion emerges only when contributors jointly concentrate on the same service. On the other hand, the weighting term $\alpha^{(s)}_{ij}$ reflects \textit{contribution prominence}. Pairs involving core contributors have greater influence, whereas peripheral contributors contribute less. This prevents cohesion from being dominated by minor or sporadic activity.

Together, these design choices ensure that PTC captures both the \textit{extent} and the \textit{uniformity} of contributor focus. The resulting $PTC_s \in [0,1]$, where higher values indicate stronger organizational cohesion. Accordingly, PTC should be interpreted as contribution-based ownership concentration, not as a direct measure of communication quality, formal team membership, or managerial task assignment.


\subsection{Measuring Organizational Coupling}

To complement cohesion, we measure cross-service interdependence using Organizational Coupling (OC) \cite{li2023evaluating}. Following prior work, the coupling between two services $M_a$ and $M_b$ is defined as:


\begin{equation}
OC(M_a, M_b) = \sum_{i=1}^{p}\left[\left(\frac{2 C_{i,a} C_{i,b}}{C_{i,a} + C_{i,b}}\right)\times S_{D_i}(M_a, M_b)\right]
\end{equation}

where $C_{i,a}$ and $C_{i,b}$ denote contributions of developer $i$ to the two services, and $S_{D_i}(M_a, M_b)$ captures the frequency with which the developer alternates between them.





To ensure comparability, we normalize the coupling as follows.

\begin{equation}
NOC(M_a, M_b) = \frac{\sum_{i=1}^{p}\left[\left(\frac{2 C_{i,a} C_{i,b}}{C_{i,a} + C_{i,b}}\right)\times S_{D_i}(M_a, M_b)
\right]}{\sum_{i=1}^{p}\left(\frac{2 C_{i,a} C_{i,b}}{C_{i,a} + C_{i,b}}\right)}
\end{equation}

This formulation ensures that $NOC(M_a, M_b) \in [0,1]$, where 1 represents maximal coupling.



We then compute \textbf{Average Organizational Coupling (AOC)} for each service:

\begin{equation}
AOC(M_a) = \frac{1}{N - 1} \sum_{b \neq a} NOC(M_a, M_b)
\end{equation}

where $N$ is the total number of microservices. Herein, AOC is used as a normalized service-level aggregation of prior OC rather than as a wholly new coupling construct.



\paragraph{Summary}

The proposed metrics capture complementary dimensions of the organizational structure. \textbf{PTC (cohesion)} captures the concentration and balance of contributor activity within a service while \textbf{AOC (coupling)} captures the extent of contributor overlap across services. Both are derived from version-control data and normalized to $[0,1]$, enabling direct comparison across services, time periods, and systems.


\section{Study Design}
\label{sec:studydesign}

To evaluate the proposed metrics, we adopt a two-level empirical design that combines a longitudinal case study with cross-system replication. 

\subsection{Overview}

Our study consists of two complementary components: 1) \textbf{Longitudinal case study:} We analyze the Spinnaker microservice platform to examine how cohesion and coupling evolve over time under real organizational and architectural changes. 2) \textbf{Cross-system replication:} We apply the same analysis pipeline to multiple open-source microservice systems to evaluate whether observed patterns generalize beyond a single case. The two research questions introduced in Section \ref{sec:introduction} guide the metric definition, longitudinal analysis, and cross-system replication. This design enables both \textit{depth} (through longitudinal analysis) and \textit{breadth} (through replication), mitigating threats to external validity.


\subsection{Data Collection and Processing}


For each project, we extracted commit histories using the GitHub REST API and mapped changed files to services using project-specific path rules. 
Files outside identifiable service boundaries, e.g., documentation, shared configuration, or build metadata, were excluded unless unambiguously assignable. Developer identities were normalized by email/name aliases, and automated commits were filtered using author/login patterns.

We then aggregated the commits at the developer-service level and normalized them into fixed time windows. For services with fewer than two contributors in a window, PTC is undefined because no contributor pair exists; such observations are excluded from pairwise correlation analyses. For Spinnaker’s repository restructuring, we maintained stable logical service names across pre- and post-restructuring mappings. These steps produce a unified dataset from which PTC and AOC are computed.


\subsection{Longitudinal Case Study: Spinnaker}

We use Spinnaker as a representative large-scale microservice system to illustrate how the proposed metrics behave under real-world evolution. Spinnaker is an open-source continuous delivery platform composed of 12 loosely coupled microservices. Its long development history, active contributor base, and documented organizational transitions make it well suited for longitudinal analysis. 

We analyze commit activity from 2017 to 2025, dividing the data into six-month windows. This allows us to observe how cohesion (PTC) and coupling (AOC) evolve in response to major events, such as governance transition to the Continuous Delivery Foundation (CDF), service lifecycle changes (e.g., deprecation) and repository restructuring (polyrepo to monorepo) \cite{spinnaker2019,spinnaker_governance_2020}. We collect commit data for 12 Spinnaker microservices from 2017 to 2025. After filtering automated commits, the dataset contains 16,911 commits, 712 developers, and 87,642 file changes across services.


The Spinnaker case provides detailed insight into how organizational dynamics are reflected in contribution patterns. However, it serves primarily as an \textit{illustrative example} rather than as the sole basis for conclusions.

\subsection{Cross-System Replication}

To assess generalizability, we replicate the analysis across six additional open-source microservice systems spanning different domains and scales. These projects vary in size, contributor base, and architectural organization, providing a heterogeneous evaluation setting. The selected microservice-based systems are listed in Table \ref{tab:project_activity}.


\begin{table}[!ht]
\centering
\begin{tabular}{lrrrrrr}
\toprule
\textbf{\#}&\textbf{project} & \textbf{\#Devs} & \textbf{\#commits} & \textbf{Start} & \textbf{End} & \textbf{Days} \\
\midrule
P1 & 1-Platform/one-platform & 25 & 908 & 2020-01-04 & 2025-11-11 & 2137 \\
P2 & camunda/camunda & 428 & 73870 & 2016-03-02 & 2026-03-12 & 3662 \\
P3 & dotnet/eShop-lineage & 270 & 4446 & 2016-09-06 & 2026-01-06 & 3409 \\
P4 & geoserver/geoserver-cloud & 27 & 2034 & 2020-07-09 & 2026-01-30 & 2031 \\
P5 & ovh/cds & 126 & 5992 & 2016-10-11 & 2026-01-20 & 3388 \\
P6 & taskcluster/taskcluster & 269 & 22390 & 2013-12-31 & 2026-01-16 & 4399 \\
\bottomrule
\end{tabular}
\caption{Summary of replicated projects}
\label{tab:project_activity}
\end{table} 

For each system, we apply the same analysis pipeline: 1) Extract commit histories and map contributions to services, 2) Compute developer focus and contribution weights, 3) Calculate PTC for each service, 4) Compute AOC based on cross-service contributions, and 5) Analyze the relationship between cohesion and coupling. This replication enables direct comparison across systems and ensures that findings are not artifacts of a single project.

\subsection{Analysis Strategy}

We analyze the data in three dimensions: 1) \textbf{Longitudinal trends:} evolution of cohesion and coupling over time (Spinnaker), 2) \textbf{Cross-system variation:} distribution of PTC and AOC across projects, and 3) \textbf{Metric relationship:} correlation between cohesion and coupling. In addition, we perform robustness analyses (Section~\ref{sec:robustness}) to evaluate the sensitivity of results to data perturbations, contributor effects, metric design, and temporal aggregation.


\paragraph{Summary}

By combining a detailed longitudinal case study with cross-system replication, our study design provides both contextual understanding and generalizable evidence. This approach enables us to evaluate whether organizational cohesion and coupling represent consistent properties of microservice development, rather than project-specific phenomena.


\section{Results}
\label{sec:results}

This section presents the empirical findings of our study. We first summarize the longitudinal case study of Spinnaker to illustrate how the proposed metrics behave under real-world evolution. We then present the cross-system replication results, which form the primary basis for our conclusions.

\subsection{Longitudinal Case Study: Spinnaker}

\begin{figure*}[!ht]
    \centering
    \begin{subfigure}[t]{0.48\textwidth}
        \centering
        \includegraphics[width=\linewidth]{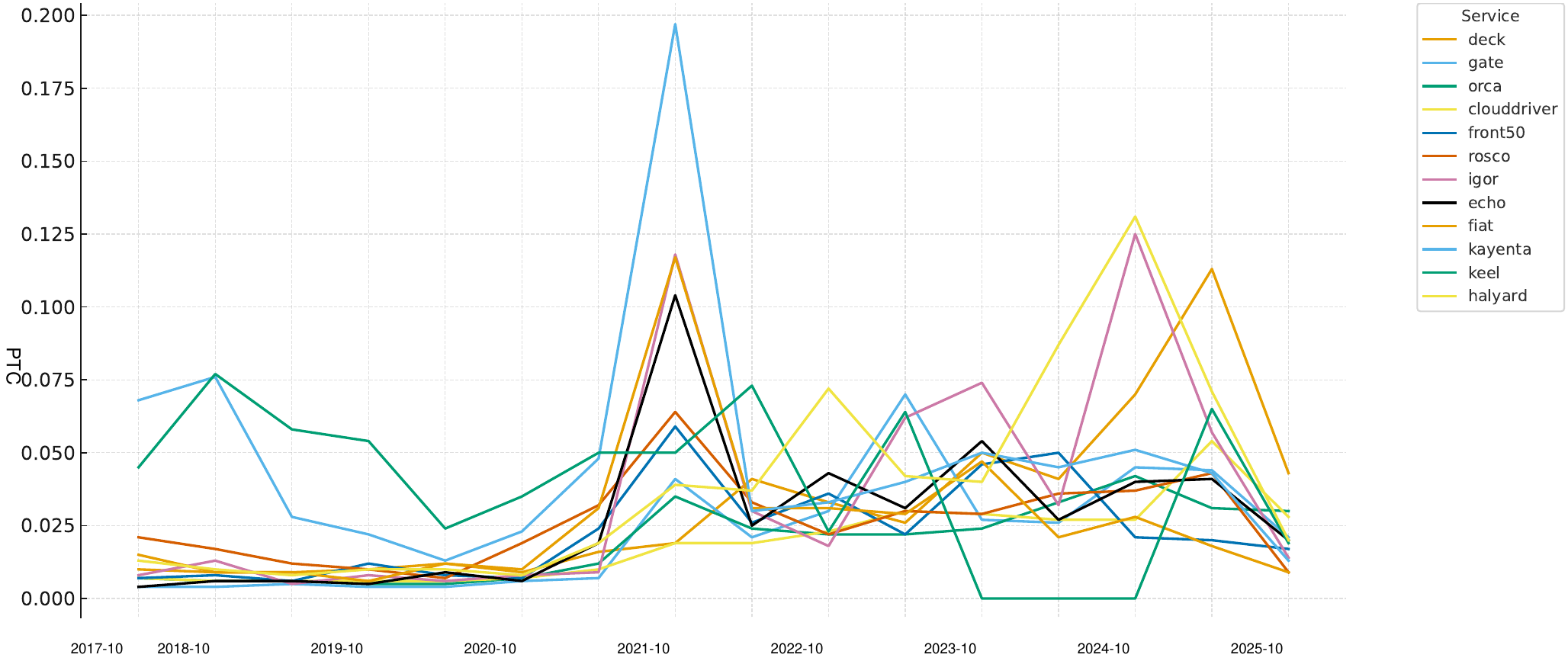}
        \caption{PTC evolution}
        \label{fig:spinnaker-PTC-evolution}
    \end{subfigure}
    \hfill
    \begin{subfigure}[t]{0.48\textwidth}
        \centering
        \includegraphics[width=\linewidth]{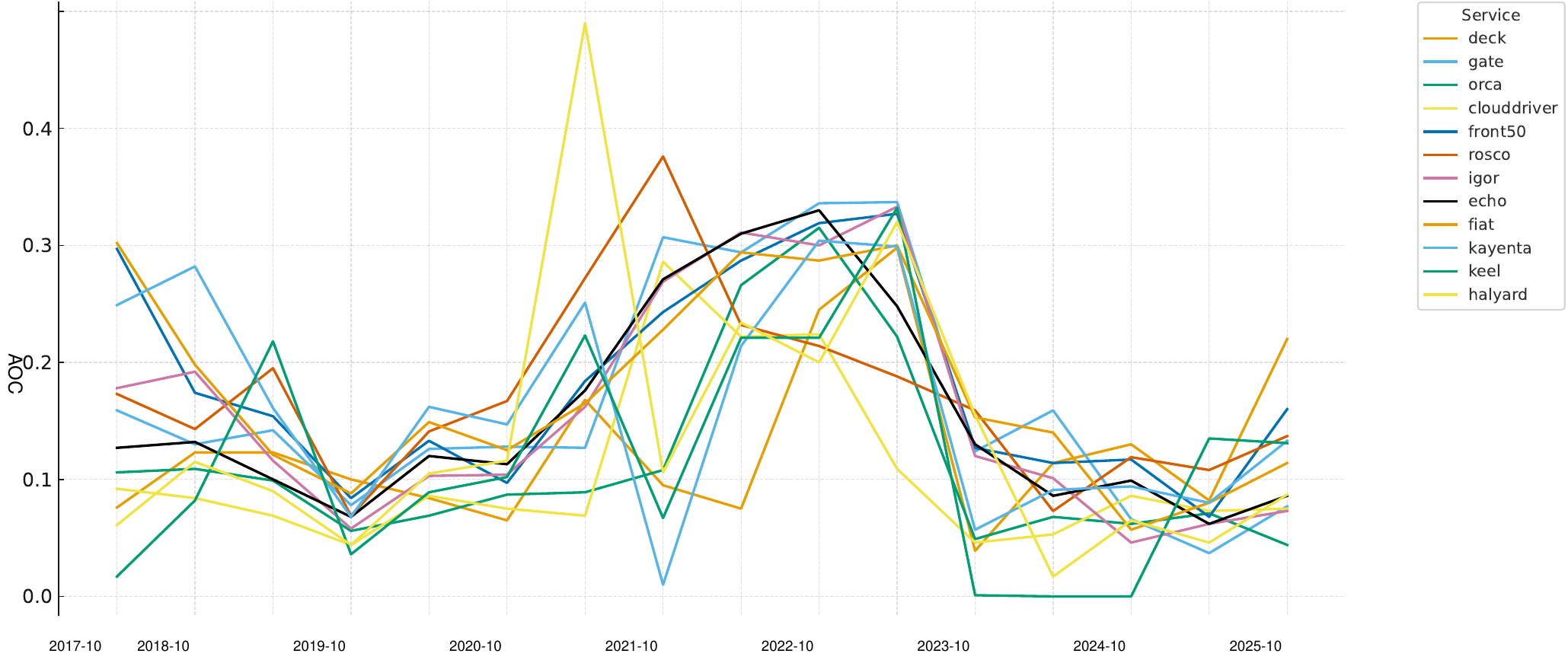}
        \caption{AOC evolution}
        \label{fig:spinnaker-aoc-evolution}
    \end{subfigure}
    \caption{Evolution of organizational cohesion (PTC) and coupling (AOC) for 12 Spinnaker microservices across 6-month windows from 10.2017 to 10.2025.}
    \label{fig:spinnaker-evolution}
\end{figure*}

We analyze the evolution of organizational cohesion (PTC) and coupling (AOC) in Spinnaker across six-month windows from 2017 to 2025 (shown in Fig. \ref{fig:spinnaker-evolution}).

Overall, PTC reflects meaningful organizational dynamics. Early periods exhibit relatively low cohesion in core services and higher cohesion in specialized services, indicating differences in ownership concentration. During governance stabilization (around 2020–2021), cohesion increases across services, suggesting more clearly defined responsibilities. In later periods, cohesion decreases as community participation broadens and cross-service contributions increase. A sharp drop is observed during the transition to a monorepo, where contributors interact across multiple services within the same repository.

The AOC exhibits complementary behavior. The early stages show relatively high coupling due to shared contributors across services. As the system matures, coupling decreases, reflecting clearer ownership boundaries. However, coupling increases again during periods of organizational change and increases sharply in the monorepo phase, where cross-service contributions become more frequent.


Across all periods in the Spinnaker case, the relationship between PTC and AOC remains weak. The longitudinal Spinnaker-only correlation is small (Pearson $r \approx 0.10$, Spearman $\rho \approx 0.15$), indicating no strong linear or monotonic association within this case. We interpret this result as case-specific evidence that cohesion and coupling capture different temporal patterns in Spinnaker, rather than as evidence of statistical independence.

These results demonstrate that the proposed metrics capture meaningful temporal patterns. However, conclusions based on a single system may be limited; therefore, we turn to cross-system replication.

\subsection{Cross-System Replication}

To assess generalizability, we analyze six additional microservice systems using the same methodology. Each observation corresponds to a service within a project.

\subsubsection{Organizational Cohesion Across Systems}

\begin{figure}[!ht]
    \centering
    \begin{subfigure}[t]{0.48\linewidth}
        \centering
        \includegraphics[width=\linewidth]{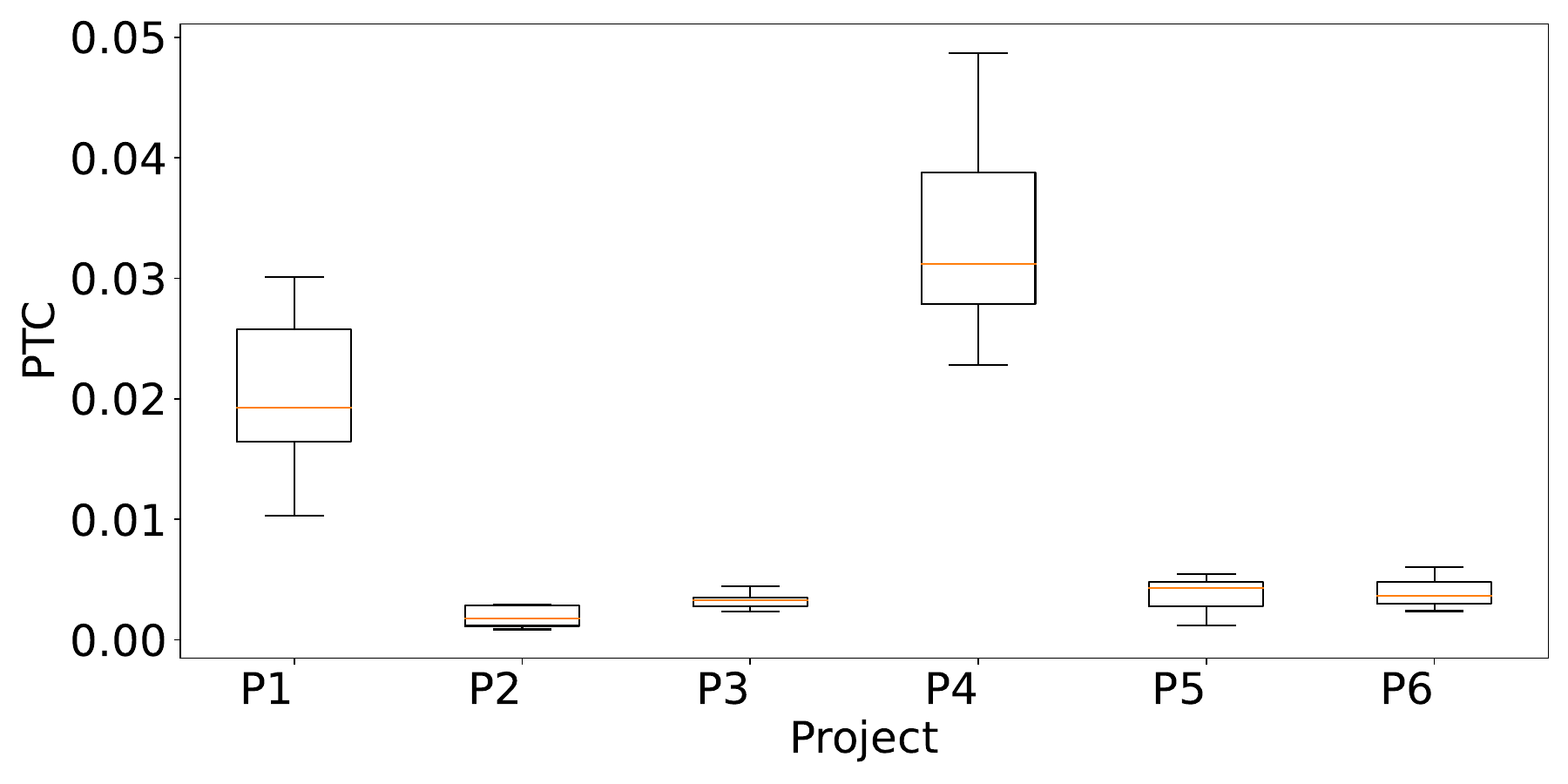}
        \caption{PTC distribution by project}
        \label{fig:ptcdistribution}
    \end{subfigure}
    \hfill
    \begin{subfigure}[t]{0.48\linewidth}
        \centering
        \includegraphics[width=\linewidth]{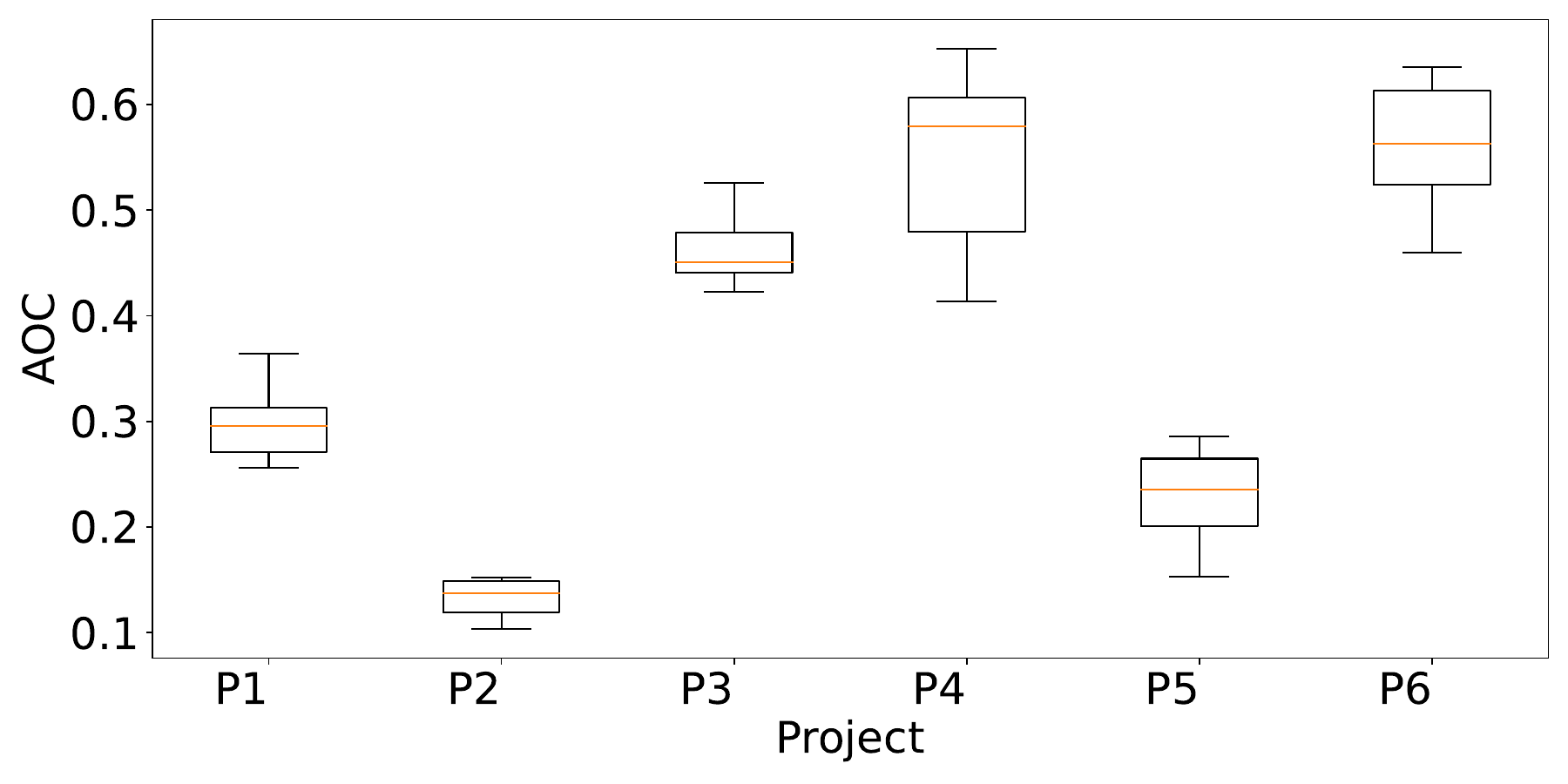}
        \caption{AOC distribution by project}
        \label{fig:aocdistribution}
    \end{subfigure}
    \caption{Distribution of organizational cohesion (PTC) and coupling (AOC) across projects}
    \label{fig:ptc_aoc_distribution}
\end{figure}

Fig. \ref{fig:ptcdistribution} shows the distribution of PTC values across projects. Cohesion varies moderately between systems, with some projects showing higher median values, indicating more concentrated contributor activity, and others showing lower values, reflecting more distributed participation. Despite these differences, cohesion values remain within a relatively bounded range across projects. This suggests that organizational focus is a stable property of microservice systems, even in varying development contexts.

Although PTC is bounded in $[0,1]$, the empirical values are small because the pairwise focus metric averages are weighted by the prominence of the contribution. Therefore, we interpret PTC comparatively rather than as an absolute percentage. E.g., within the same project, a service around 0.04 reflects more concentrated and balanced ownership than one around 0.01, while low values indicate more diffuse or cross-service contribution patterns.

\subsubsection{Organizational Coupling Across Systems}

Fig. \ref{fig:aocdistribution} presents the distribution of the AOC values. In contrast to cohesion, coupling exhibits substantial variability across systems. Some projects show high coupling, indicating strong cross-service contributor overlap, while others display low coupling, reflecting clearer separation of service ownership. This variation suggests that coupling is highly context-dependent, influenced by factors such as architectural design, contributor practices, and repository organization.

\subsubsection{Relationship Between Cohesion and Coupling}

\begin{figure}[!ht]
    \centering
    \includegraphics[width=0.9\linewidth]{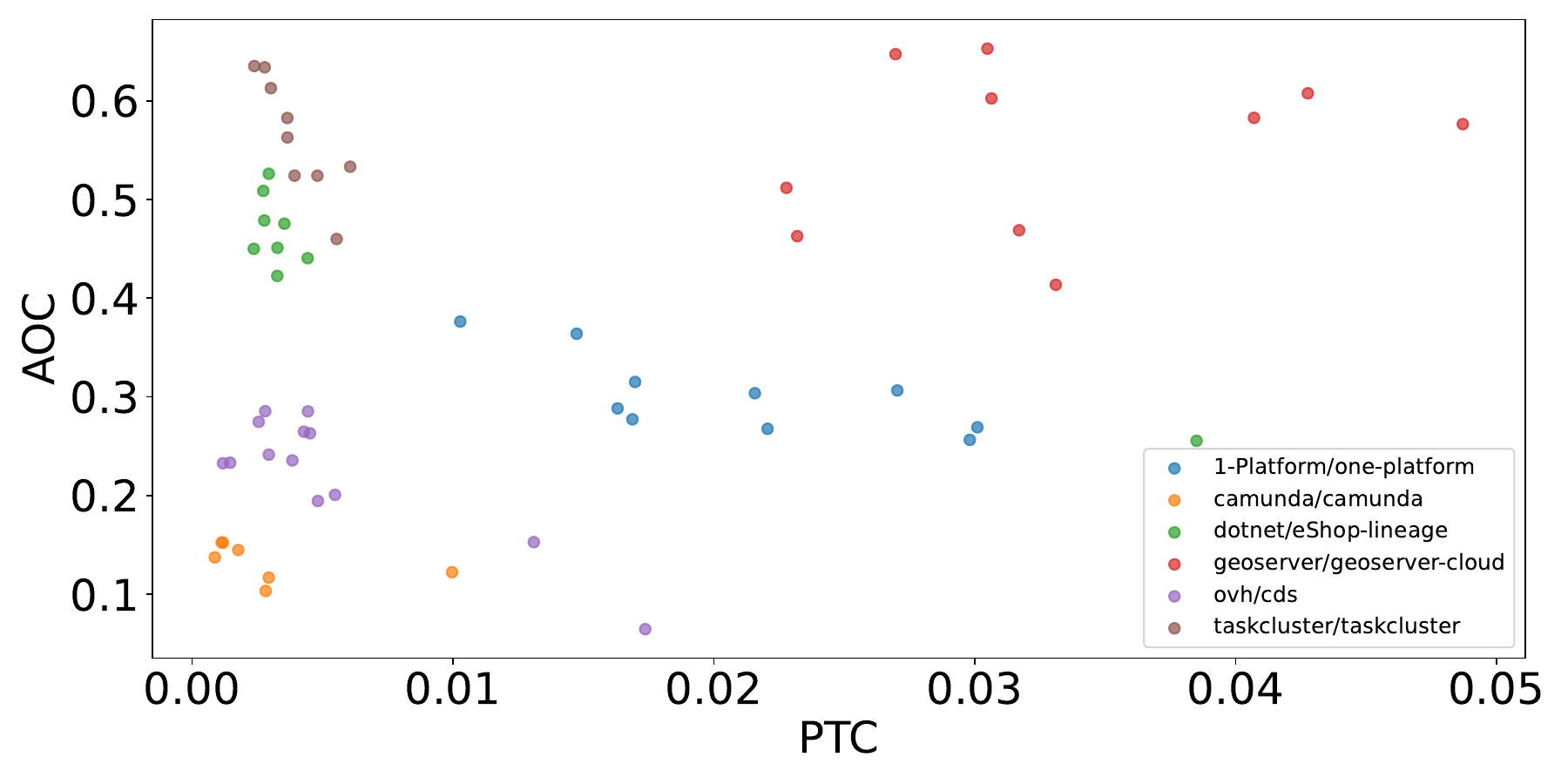}
    \caption{PTC vs AOC across all projects}
    \label{fig:ptcaocscatter}
\end{figure}

Figure~\ref{fig:ptcaocscatter} shows the relationship between PTC and AOC across all services and projects. No clear global trend is observed. Instead, services form project-specific clusters, and both high and low cohesion values are associated with a wide range of coupling levels.

This pattern is consistent with the Spinnaker case study in showing no strong global trend. However, the pooled cross-system association differs from the Spinnaker-only estimate in both sign and magnitude. In the pooled baseline analysis reported later in Table~\ref{tab:robustness-summary}, the Pearson correlation is weakly negative ($r=-0.233$, 95\% CI $[-0.30,-0.16]$, $p<0.001$), while the rank correlation is close to zero ($\rho=-0.072$). Thus, the evidence does not show the absence of any relationship; rather, it indicates a small and aggregation-sensitive association between PTC and AOC.

\subsubsection{Summary of Replication Findings}

The replication demonstrates that: 1) Organizational cohesion varies moderately across systems, while coupling varies substantially; 2) Microservice systems exhibit diverse organizational structures, ranging from tightly coupled to highly decoupled ecosystems; 3) The relationship between cohesion and coupling is consistently weak across projects.


These results suggest that cohesion and coupling should be analyzed as separate organizational dimensions rather than as opposite ends of a single continuum.
\section{Robustness Analysis}
\label{sec:robustness}


To assess the reliability of our findings, we evaluate robustness along four dimensions: data perturbations, contributor effects, metric formulation, and temporal aggregation.

\subsection{Data Robustness}





Regarding the effect of large commits, we remove the top 5\% of commits by size (P95 threshold) to exclude large refactorings and bulk changes. PTC and AOC distributions remain stable, with only minor variance reduction. The correlation between cohesion and coupling remains weak. In summary, the result is robust to commit size.





Furthermore, regarding the effect of dominant contributors, we remove the top 5\% of contributors by activity to assess the impact of core maintainers. Absolute values decrease slightly, but distributions and correlations remain consistent. In summary, the result is not driven by dominant contributors.

\subsection{Metric Robustness}




We test alternative formulations of PTC by varying connection intensity, weighting, and normalization. Although absolute values shift slightly, relative rankings and correlations remain stable across variants. In summary, the findings are not sensitive to the metric design.

\subsection{Cross-Condition Stability}

Across all robustness conditions described above, three consistent properties are observed: 1) The distributions of PTC and AOC remain stable under data filtering and metric variation; 2) The relative ordering of services and projects is preserved; 3) The correlation between cohesion and coupling is consistently weak. These results indicate that the findings are robust to both data-level noise and metric formulation choices.


\subsection{Temporal Sensitivity Analysis}






We compare overlapping windows (baseline) with non-overlapping 180-day windows. The latter increases variance and reduces sample size, leading to more variable correlations. However, no consistent relationship emerges between cohesion and coupling.

Correlation remains weak overall, though more unstable under non-overlapping windows (e.g., $r=0.259$, 95\% CI $[0.00,0.48]$, $p\approx0.05$). The wider confidence interval indicates reduced precision due to the smaller sample size and the absence of temporal smoothing. Therefore, the temporal analysis does not support a stable positive or negative relationship between PTC and AOC; instead, it shows that the estimated association is sensitive to temporal aggregation.



\subsection{Overall Robustness Interpretation}

As shown in Table~\ref{tab:robustness-summary}, the non-overlapping configuration yields a wider confidence interval and borderline statistical significance, reflecting increased variability due to reduced sample size and the absence of temporal smoothing.

\begin{table}[!ht]
\centering
\caption{Robustness summary of PTC-AOC correlation across conditions.}
\label{tab:robustness-summary}
\begin{tabular}{lrrrrrrr}
\toprule
Condition & $n$ & $r$ & 95\% CI ($r$) & $p$-value & $\rho$ & $\Delta r$ & $\Delta \rho$ \\
\midrule
Baseline 
& 745 
& -0.233 
& [-0.30, -0.16] 
& $<0.001$ 
& -0.072 
& 0.000 
& 0.000 \\

Remove large commits (P95) 
& 739 
& -0.226 
& [-0.29, -0.15] 
& $<0.001$ 
& -0.063 
& 0.007 
& 0.008 \\

Remove top contributors (top 5\%) 
& 662 
& -0.171 
& [-0.24, -0.10] 
& $<0.001$ 
& 0.019 
& 0.063 
& 0.090 \\

Non-overlapping windows (180d/180d) 
& 58 
& 0.259 
& [0.00, 0.48] 
& 0.05 
& 0.267 
& 0.492 
& 0.339 \\

\bottomrule
\end{tabular}
\end{table}


Taken together, the robustness analyses demonstrate that: 1) the results are not sensitive to large commits or extreme contribution patterns; 2) the findings are not driven by a small number of highly active contributors; 3) the observed association remains small under alternative metric formulations; and 4) the estimated correlation varies under temporal re-aggregation. The baseline pooled analysis shows a weak but statistically significant negative Pearson correlation, 
whereas Spearman's $\rho=-0.072$ indicates almost no monotonic association. Therefore, the results should be interpreted as evidence that PTC and AOC are only weakly associated and provide non-redundant information, not as evidence that they are statistically independent.


These findings provide strong evidence that the observed relationship between organizational cohesion and coupling reflects a structural property of developer behavior in microservice systems, rather than a consequence of measurement artifacts or methodological choices.


In summary, the robustness analysis confirms that the proposed metrics and empirical findings are stable under several validation conditions. In particular, the small magnitude of the PTC-AOC association persists under data perturbations and metric variations, while temporal re-aggregation affects the estimated sign and magnitude. This supports the conclusion that cohesion and coupling should be analyzed as distinct organizational dimensions, while avoiding the stronger claim that they are statistically independent.
\section{Discussion}
\label{sec:discussion}

This study introduces Pairwise Team Cohesion (PTC) as a quantitative measure of organizational cohesion in microservice systems and evaluates it through a longitudinal case study and cross-system replication. The results consistently show that PTC captures meaningful patterns of contributor focus and that cohesion and coupling represent distinct dimensions of socio-technical organization. We interpret PTC and AOC as behavioral proxies for contribution-based ownership concentration and cross-service developer overlap. Longitudinal patterns, robustness checks, and cross-system replication support their stability, but not full construct validation against formal team assignments or communication data.

\subsection{Organizational Cohesion as Behavioral Alignment (RQ1)}

Across both the Spinnaker case and replicated systems, PTC reflects the concentration and stability of contributor activity within service boundaries. Cohesion increases during periods of stabilized ownership and decreases when participation becomes more distributed, indicating sensitivity to changes in team structure and coordination. These patterns suggest that organizational cohesion can be operationalized as behavioral alignment between contributors and service boundaries, observable through contribution distributions. Although PTC does not capture communication or intent, it provides a scalable proxy for ownership concentration and participation balance. In this sense, cohesion reflects how consistently contributors focus their work within a service, rather than how they coordinate internally. 

\subsection{Cohesion and Coupling as Distinct Dimensions (RQ2)}

A central finding of this study is that the relationship between cohesion (PTC) and coupling (AOC) is weak and aggregation-sensitive. In the longitudinal Spinnaker case, the association is small and positive (Pearson $r \approx 0.10$, Spearman $\rho \approx 0.15$). In the pooled cross-system baseline analysis, the Pearson correlation is weakly negative ($r=-0.233$, 95\% CI $[-0.30,-0.16]$, $p<0.001$), while the Spearman correlation is close to zero ($\rho=-0.072$). These results do not establish statistical independence. Instead, they indicate that PTC and AOC provide non-redundant views of organizational structure: cohesion captures intra-service concentration of effort, whereas coupling reflects cross-service contributor overlap. This reframes the classical "high cohesion, low coupling" principle at the organizational level as a two-dimensional diagnostic space rather than a single inverse relationship. This interpretation is also consistent with recent evidence that organizational coupling can precede architectural degradation signals, suggesting that AOC should be read not only as a coordination indicator, but also as a potential early warning signal for architectural risk \cite{mani2026organizational}. 



    


\subsection{Implication}

The weak and aggregation-sensitive association between cohesion and coupling has direct implications for the management of microservice organizations. PTC and AOC provide complementary diagnostic signals. PTC (cohesion) helps identify ownership concentration, detect fragmentation of responsibility, and monitor the effects of organizational changes. AOC (coupling) captures cross-team dependencies and coordination load arising from shared contributors.


Because these dimensions are non-redundant and only weakly associated, relying on a single metric can be misleading. For example, a highly cohesive service can require substantial cross-team coordination, while a loosely coupled system may suffer from fragmented ownership. Effective organizational design therefore requires joint consideration of cohesion and coupling, rather than optimizing for one in isolation.

\subsection{Limitation and Future Work}

PTC is derived from version-control data and captures observable contribution behavior rather than the full spectrum of collaboration. It does not account for communication, code review interactions, or informal coordination mechanisms, and may be influenced by development practices such as commit granularity or repository structure.

Future work should extend this approach by integrating additional data sources (e.g., pull requests, issue discussions, communication logs) to better capture coordination dynamics. Another direction is to examine how different cohesion–coupling configurations relate to software outcomes such as quality, delivery performance, and system evolution, enabling more prescriptive guidance for organizational design. 
This study does not test whether PTC/AOC align with structural service cohesion or technical coupling of the same services. Therefore, our conclusions are limited to the contribution-based organizational structure. Future work should compare PTC/AOC with service-level structural cohesion and technical coupling to determine whether organizational and architectural modularity evolve together or diverge. Beyond passive diagnosis, recent work also suggests that organizational cohesion and coupling metrics can be integrated into gamified architectural governance mechanisms to actively encourage service-boundary-respecting developer behavior \cite{li2026gamifying}.



\section{Threats to Validity}
\label{sec:threat}

We discuss threats to validity following the classification of Wohlin et al.\cite{wohlin2012experimentation}, covering internal, external, construct, and conclusion validity.


Regarding internal validity, this study is observational and does not establish causal relationships between organizational events and changes in cohesion or coupling. 
Although variations in PTC and AOC align with documented milestones (e.g., governance stabilization and repository restructuring), they may also be influenced by unobserved factors such as contributor turnover, policy changes, or CI/CD workflow evolution.
To mitigate this risk, we interpret results conservatively and triangulate metric trends with publicly available project artifacts (e.g., governance documentation and release notes). Both metrics are computed from the same dataset and time windows, ensuring consistency in measurement. Nevertheless, findings should be interpreted as correlational rather than causal.


Regarding external validity, the primary case study (Spinnaker) represents a large, mature open-source system with hybrid governance, which may limit generalizability to smaller or purely industrial projects. 
We address this through replication across multiple microservice systems with different sizes and contribution structures. Therefore, generalization should focus on relative patterns rather than fixed numeric thresholds.


Regarding construct validity, PTC and AOC capture version-control activity rather than formal team structure or the full spectrum of collaboration. 
They do not account for informal coordination mechanisms, e.g., discussions, code reviews, or meetings, and commit activity itself may be influenced by workflow practices, tooling, or repository structure. 
Hence, we reduce bias by filtering identifiable automated commits and testing sensitivity to large commits, dominant contributors, and metric variants. Thus, PTC should be interpreted as contribution-based cohesion, not complete team cohesion. 


Regarding conclusion validity, the study uses a longitudinal multi-project dataset and reports both Pearson and Spearman correlations. 
Robustness analyses further confirm that the weak relationship between cohesion and coupling persists under data filtering, metric variations, and temporal aggregation strategies. 
However, non-overlapping windows yield smaller samples and wider confidence intervals, so exact effect sizes remain sensitive to temporal granularity.
\section{Conclusion}
\label{sec:conclusion}

This study introduces Pairwise Team Cohesion (PTC) as a quantitative metric for assessing organizational cohesion in microservice systems and evaluates it through a longitudinal case study combined with cross-system replication and robustness analyses. The results show that PTC captures the concentration and stability of contributor activity within service boundaries and responds systematically to organizational and architectural changes. Across the analyzed systems and robustness conditions, the relationship between cohesion (PTC) and organizational coupling (AOC) remains weak, although its sign and magnitude vary with the level of aggregation. The Spinnaker-only longitudinal analysis shows a small positive association, whereas the pooled baseline analysis shows a weak but statistically significant negative Pearson correlation and an almost zero Spearman correlation. These results indicate that PTC and AOC capture distinct, non-redundant aspects of socio-technical organization rather than a single inverse cohesion--coupling trade-off. This reframes the classical "high cohesion, low coupling" principle as a two-dimensional diagnostic space and highlights the need to evaluate internal team focus and cross-service interaction separately.

\bibliographystyle{splncs04}
\bibliography{bib}
\end{document}